\documentclass[%
reprint,
superscriptaddress,
showpacs,
 amsmath,amssymb,
 aps,
prb,
twocolumn,
oneside,
floatfix,
]{revtex4-1}
\usepackage{graphicx, subfigure}
\usepackage{float}
\usepackage{dcolumn}
\usepackage{bm}
\usepackage[utf8]{inputenc}
\usepackage{relsize}
\usepackage{amssymb,bbm}
\usepackage{makecell}
\usepackage{braket}
\usepackage{xcolor}
\usepackage{color}
\usepackage{soul}

\newcommand{\si}{\sigma}
\newcommand{\eps}{\epsilon}
\newcommand{\+}{^{\dagger}}

\newcommand{\ua}{\uparrow}
\newcommand{\da}{\downarrow}

\def\bra#1{\mathinner{\langle{#1}|}}
\def\ket#1{\mathinner{|{#1}\rangle}}
\def\braket#1{\mathinner{\langle{#1}\rangle}}

\def\Ket#1{\left|#1\right\rangle}

\newcommand{\blue}[1]{{#1}}
\newcommand{\bblue}[1]{{#1}} 

\newcommand{\sbb}[1]{#1}
\renewcommand{\st}[1]{} 

\newcommand{\etal}{\textit{et al.\/}}
\newcommand{\ie}{i.\,e.}

\newcommand{\tc}{\textcolor}

\newcommand* \dzq{\ensuremath{\text{d}_{z^2}~}}
\newcommand* \dxq{\ensuremath{\text{d}_{x^2-y^2}~}}
\newcommand* \dxz{\ensuremath{\text{d}_{xz}~}}
\newcommand* \dyz{\ensuremath{\text{d}_{yz}~}}
\newcommand* \dxy{\ensuremath{\text{d}_{xy}~}}

\newcommand* \dzqX{\ensuremath{\text{d}_{z^2}}}
\newcommand* \dxqX{\ensuremath{\text{d}_{x^2-y^2}}}

\graphicspath{{./fig_plots/}}
\usepackage[outdir=./]{epstopdf}

\begin{document}

\preprint{APS/123-QED}

\title{Role of charge transfer in hybridization-induced spin transition in metal-organic molecules}
\author{Jakob Steinbauer}
\email{jakob.steinbauer@polytechnique.edu}
\affiliation{CPHT, CNRS, Ecole Polytechnique, Institut Polytechnique de Paris, F-91128 Palaiseau, France}
\author{Silke Biermann}%
\affiliation{CPHT, CNRS, Ecole Polytechnique, Institut Polytechnique de Paris, F-91128 Palaiseau, France}
\affiliation{Coll\`{e}ge de France, 11 place Marcelin Berthelot, 75005 Paris, France}
\affiliation{European Theoretical Spectroscopy Facility, Europe}
\author{Sumanta Bhandary}
\email{sumanta.bhandary@polytechnique.edu}
\affiliation{CPHT, CNRS, Ecole Polytechnique, Institut Polytechnique de Paris, F-91128 Palaiseau, France}

\date{\today}

\begin{abstract}
The spin-crossover in organometallic molecules constitutes one of the most promising routes towards the realization of molecular spintronic devices. In this article, we explore the hybridization-induced spin-crossover in metal-organic complexes. We propose a minimal many-body model that captures the essence of the spin-state switching \st{in a generic parameter space}, thus providing  \bblue{insight into the} underlying physics. Combining the model with density functional theory (DFT), we then study the spin-crossover in isomeric structures of Ni-porphyrin (Ni-TPP). We show that metal-ligand charge transfer plays a crucial role in the determination of the spin-state in Ni-TPP. Finally, we propose a spin-crossover mechanism based on mechanical strain, which does not require a switch between isomeric structures. 

\end{abstract}

\pacs{Valid PACS appear here}
\maketitle

\section{Introduction}
Molecular spintronics\cite{spintronics_bogani,spintronics_rocha,spintronics_sanvito} based on single molecules inherits a major advantage over its bulk counterpart: Magnetism in a solid-state context is a co-operative phenomenon involving a large number of neighboring atoms, while molecular magnetism can emerge within a single site in a molecular system. The prospective gain in miniaturization using organic molecules is, therefore, enormous. In addition, molecular spintronic devices offer the possibility of efficient information processing and low power dissipation.
\sbb{These promising perspectives have propelled the advancement of single molecule-based spintronics in the last two decades. \par
The realization of molecular spintronic devices, such as molecular-valves\cite{valve}, -switches\cite{switch_ormaza, switch_sco} or information storage devices\cite{memory_1, memory_2,memory_3}, relies on the realization of magnetic bi-stability and its controllability through the coupling to external stimuli. State-of-the-art proposals include molecules with different spin-states, magnetic couplings, magnetic anisotropy or presence/absence of Kondo resonance\cite{kondo_transistor}. Considering  control and reversibility, systems exhibiting molecular spin-crossovers (SCO) are among the most promising. Indeed, SCO can be controlled by multiple external stimuli, such as temperature, light, pressure, electric fields, ligand adsorption or mechanical strain\cite{sco_1,sco_2,mol_sco,sumanta_prl,switch_ormaza,C2CP40111H}. 
%
%
Organic molecular complexes, hosting transition metal ions  such as Fe$^{2+}$, Fe$^{3+}$, Co$^{2+}$, Ni$^{2+}$, Mn$^{2+}$, Mn$^{3+}$ often exhibit spin state switching due to a subtle balance between ligand field and spin-pairing energy. An efficient way to access the spin state is by modifying the structure, which in turn changes the ligand field. \par 
In 2011, light-induced excited spin state trapping in a thin film of iron molecular complexes has been observed for the first time\cite{film1}; up to date, this remains one of the most effective ways to control the SCO\cite{film2}. Using temperature and light, Kuch \etal~achieved a spin-crossover in molecules adsorbed on Au\cite{kuch1} and graphite \cite{kuch2,iron2_SOC_graphite_kuch} surfaces. While these observations were achieved at low temperatures, recently also room temperature spin state switching has been observed in iron molecular complexes, both in solution\cite{sol,room_temp_science} and in the solid state\cite{solid}. Another efficient route to switch spin states is a transformation between isomeric structures, through ligand association or dissociation. Herges \etal~have shown that upon irradiation, a spin-crossover can be induced\cite{fe3,Nip_6} in porphyrin molecules with photochromic axial ligands. \par
The emergence of scanning tunneling microscopy (STM) provides unprecedented control to manipulate properties at the single atomic/molecular scale. In Fe-complexes, adsorbed on a metallic surface, Miyamachi \etal~achieved a spin-crossover by controlling the metal-molecule interaction with a STM tip\cite{stm}. Furthermore, with the aid of a STM tip, a voltage pulse can be applied, which can induce a spin-crossover in molecular complexes\cite{voltage}.}

\sbb{Describing spin-crossover phenomena theoretically is a challenging task. Different approaches, such as density functional theory (DFT) with several forms of exchange-correlation functionals\cite{sco_dft1,sco_dft2},  DFT+U \cite{sumanta_prl,sco_dftu,sumanta_surf}, DFT+many-body theory\cite{dft+dmft, sumprb}, full configuration interaction quantum Monte Carlo \cite{alvi} methods, etc. have been used in the literature. On the specific example of nickel porphyrin\cite{theo_nip}, time-dependent density functional theory techniques have been used to explicitly characterize the singlet and triplet excited states. \par
The basic mechanism underlying the various strategies of inducing SCOs is a controlled manipulation of the molecular ligand field. In this article, we explore the role of TM-ligand hybridization in organometallic molecules. In particular, we focus on the isomers of  the nickel tetraphenylporphyrin (Ni-TPP) molecule, and the way in which the hybridization determines their spin states. Due to the strong axial bond formation between metal ions and organic ligands, charge transfer might be significant, as also suggested in\cite{theo_nip}.\par 
In this paper, we demonstrate the importance of the metal-ligand charge transfer to realize a spin-crossover, by investigating the interplay between hybridization, crystal field and Coulomb interaction. To this means, we construct a minimal model which captures the relevant physics of the spin-crossover. The model is then used \bblue{for a} realistic \bblue{description of} molecular systems by importing model parameters from density functional theory (DFT) calculations. Finally, we detail a strain-assisted mechanism for spin-state switching in Ni-TPP molecules. The realization of such a mechanism gives promise to a potential integration in a mechanically controlled break junction (MCBJ)\cite{mcbj1,mcbj2,mcbj3} or scanning tunneling microscopy break junction (STM-BJ)\cite{stmbj, STM_break_junction} device. }

\par The paper is organized as follows. In section \ref{model_A}, we construct a generic many-body model, that takes into consideration the essential physics behind the spin-crossover. Subsection \ref{sec:generic_results} then describes the SCO mechanism for a set of generic parameters. In Sec. \ref{sec:real_system}, based on first-principle calculations, the model is employed to describe the SCO scenario in Ni-TPP. Section \ref{sec:conclusion} summarizes our findings.

\section{Hybridization-induced spin-crossover}\label{model}
\subsection{The model}\label{model_A}
\begin{figure}
\begin{center}
\includegraphics[width=0.49\textwidth ]{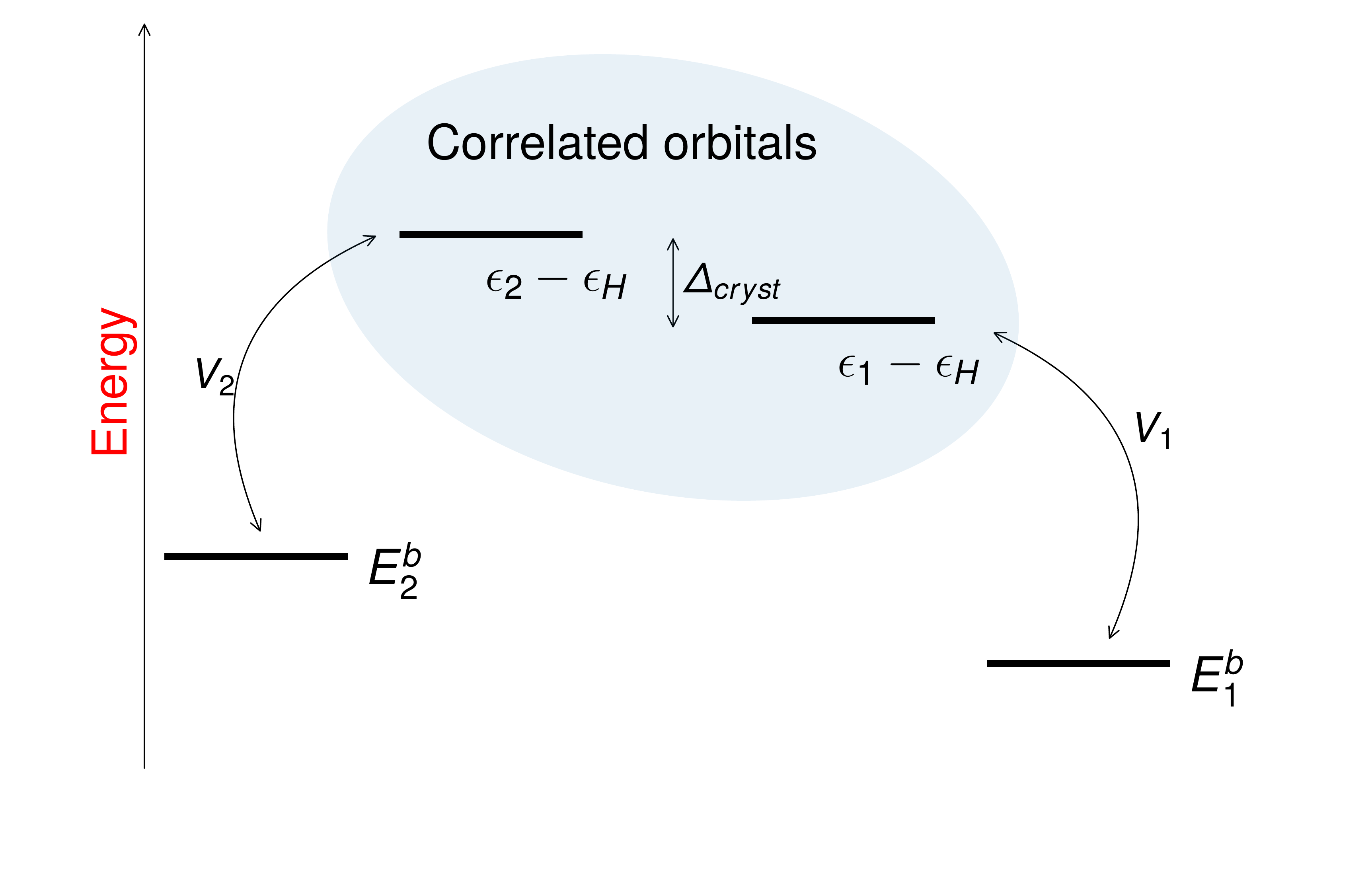}
\end{center}
\caption{Schematic illustration of our model. The system consists of two correlated orbitals with bare energies $\eps_{1}$/$\eps_{2}$, interacting via Coulomb repulsion and Hund's exchange coupling. The difference of the onsite energies of the correlated orbitals is denoted by $\Delta_{cryst} = \eps_{2}-\eps_{1}$. Each metal orbital couples to an uncorrelated ligand orbital with a hybridization strength $V_{1}$/$V_{2}$; the energy levels of the ligand orbitals are $E^{b}_{1}$/$E^{b}_{2}$. All bare energy levels will be shifted by a chemical potential defined by fixing the overall filling. }\label{fig:modelschematic}
\end{figure}

Our goal is to describe the changes in the spin state of a transition metal ion brought about by a modification of the ligand to transition metal hybridization strength. To this effect, we construct a model, in which the transition metal ion and the ligands are represented by two orbitals each. A minimal number of two orbitals is necessary to incorporate the effect of Hund's exchange, which is at the heart of any high-spin configuration. 
Our generic model for the description of the SCO, illustrated in Fig. \ref{fig:modelschematic}, 
is defined by the following Hamiltonian 
\begin{align}\label{eq:the_model}
\begin{split}
H &= \sum_{m=1,2,\si}(\epsilon_{m} - \epsilon_{H}) n_{m\si} + \sum_{m=1,2} E^{b}_{m}\sum_{\si} n^{b}_{m\si}\\ 
&+  \sum_{m=1,2}\sum_{\si} \left(V_{m} d_{m\si}^{\+}b_{m\si}+ h.c.\right)  \\
&+ U\sum_{m=1,2}n_{m\ua}n_{m\da}+ \sum_{\si\si'} (U' - \delta_{\si\si'}J)n_{1\si}n_{2\si'}\\
&-\mu\sum_{m=1,2}\sum_{\si} (n_{m\si} +n^{b}_{m\si})\text{ ,}
\end{split}
\end{align}
where $d^{\+}_{m\si} $($d_{m\si}$) create (annihilate) an electron at the correlated orbital $m$ (of energy $\epsilon_{m}$) with spin $\si$, while $b^{\+}_{m\si}$($b_{m\si}$) denote the creation (annihilation) operators of the electrons at the ligand orbitals $m$ (of energy $E^{b}_{m}$). The number operators are defined as $n_{m\si}=d^{\+}_{m\si}d_{m\si}$ and $n^{b}_{m\si} = b^{\+}_{m\si}b_{m\si}$. $V_{m}$ is the hybridization between correlated and ligand orbitals, $U$ and $J$ represent onsite Coulomb- and exchange interactions, respectively, with $U' = U-2J$. 

The chemical potential, $\mu$ fixes the overall occupation (correlated orbitals+ligands). Throughout the paper, this overall occupation will be fixed to 6. The term $\epsilon_{H}$ shifts the energy levels $\epsilon_{m}$ and is explicitly added to cancel the Hartree contribution from the interaction. In the context of our model, we shall define the crystal field splitting as the difference of the energy levels of the correlated orbitals $\Delta_{cryst} = \eps_{2}-\eps_{1}$.\linebreak
%
%
In realistic systems, external stimuli such as strain would typically change several model parameters. Nevertheless, with the present model, we can explore the whole parameter space 
spanned by crystal field strength and hybridization. 
A spin-crossover can then be realized in two ways -- either by crystal field modification, \ie~changing the relative energies of the correlated orbitals or by tuning the metal-ligand hybridization.
\par The Hilbert space spanned by Hamiltonian \eqref{eq:the_model} is of dimension $D = 4^{4} = 256$, and therefore easily treatable by means of exact diagonalization. This is the method pursued throughout this paper.
\subsection{Model parameters and observables}\label{sec:generic_results}
The quantity of central interest is the spin moment $\Braket{\vec{S}^{2}}$, defined as 
\begin{align}
\vec{S}_{tot} &= \frac{1}{2}\sum_{m=1,2} \sum_{\si\si'}\left( d^{\+}_{m\si}\vec{\si}_{\si\si'}d_{m\si'} + b^{\+}_{m\si}\vec{\si}_{\si\si'}b_{m\si'} \right)\\
\vec{S}_{corr} &= \frac{1}{2}\sum_{m=1,2} \sum_{\si\si'} d^{\+}_{m\si}\vec{\si}_{\si\si'}d_{m\si'} \text{ ,}
\end{align}
where $\vec{S}_{tot}$ and $\vec{S}_{corr}$ describe the spin moment of the total molecule and the correlated subspace only, respectively. The occupations of the correlated orbitals $n_{m} =\sum_{\si}n_{m\si}$ will provide information about the charge transfer from the ligand orbitals to the correlated orbitals. Furthermore, we consider the free energy $F = \Braket{H} - ST$, to analyze the energetics of the different spin configurations.\par
These quantities will be calculated as a function of the crystal field $\Delta_{cryst}$ and \blue{the ratio of the hybridization strengths $V_{1}/V_{2}$, for fixed $V_{2}= \sqrt{8} eV$.} \blue{Throughout the following model study, we take $U=5.14$eV and $J=0.89$eV\footnote{Even though the model depends only on the ratio of the parameters e.g. with respect to $V_{2}$, we prefer here to introduce electron Volts as a unit, to enable an easier comparison with the following chapters.}.} Furthermore, we consider the parameters $E^{b}_{1}=E^{b}_{2} = -2 eV$, while the bare e$_{g}$ levels will be set to $\epsilon_{1} = - \Delta_{cryst} - \epsilon_{H}$ and $\epsilon_{2} = + \Delta_{cryst} - \epsilon_{H}$. 
The Hartree potential, corresponding to a homogeneous charge distribution, reads $\epsilon_{H} = \frac{N}{4}(3U - 5J)$. In principle, $N$ should be the total occupation of the correlated orbitals. However, inspired by the fully localized limit double-counting of electronic structure theory\cite{Lichtenstein_LDA+U}, we rather choose the integer values $N=2$ or $N=3$.\par 
In the case without hybridization, the correlated orbitals would have an occupation of $N=2$, while for the above parameters, the ligands would be completely filled. The high-spin state would then be associated with having one electron per correlated orbital, while the low-spin state would correspond to the orbitally-polarized configuration. In this case, $\Braket{\vec{S}_{tot}^{2}} =\Braket{\vec{S}_{corr}^{2}}$, with $\Braket{\vec{S}_{corr}^{2}} = 2$ and $0$, respectively in the high-spin and low-spin configurations. 

\begin{figure}[t]
\begin{center}
\includegraphics[scale=0.42]{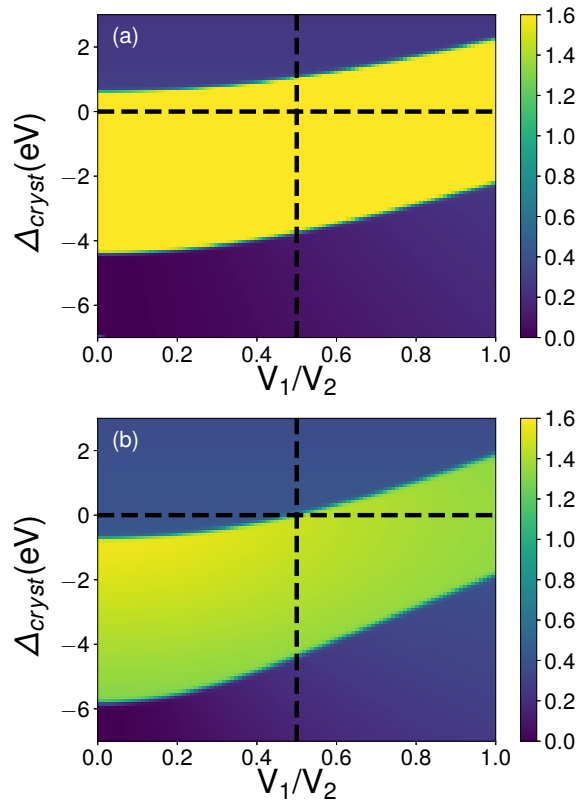}
\caption{Spin moment $\Braket{\vec{S}_{corr}^{2}}$ of the correlated orbitals as a function of the crystal field $\Delta_{cryst} = (\epsilon_{2}-\epsilon_{1})$ and the $d_{z^2}$ hybridization for $\epsilon_{H}(N=2)$ (upper panel) and $\epsilon_{H}(N=3)$ (lower panel). Fig. \ref{fig:equal_V1_CF_cut} shows different quantities for constant $V_{1}$ along the dashed line. }\label{fig:generic_phasediagram}
\end{center}
\end{figure}

\subsection{Results}\label{sec:Results_onlymodel}
Fig. \ref{fig:generic_phasediagram} shows the spin moment $\Braket{\vec{S}^{2}}$ as a function of $\Delta_{cryst}$ and $V_{1}$ for $\epsilon_{H}(N=2)$ (upper panel (a)) and $\epsilon_{H}(N=3)$ (lower panel (b)). Both figures exhibit two low-spin regions (blue), \blue{separated by a band-like high-spin region (yellow) of a width that decreases from about $\sim 5.5J$ to  $\sim4.5J$ upon increasing $V_{1}/V_{2}$.}  Qualitatively, the shape of the high-spin region is easily explained: In the \emph{atomic limit}, in which both hybridizations vanish ($V_{1}=V_{2} = 0$), a high-spin to low-spin transition would be induced as soon as $|\epsilon_{2}-\epsilon_{1}| = |\Delta_{cryst}| > 3J $, therefore resulting in a width of $6J$. While in the absence of hybridizations this region would be symmetric around $\Delta_{cryst}=0$, a finite $V_{2}>0 $ will lead to two molecular orbitals; a bonding orbital with its energy below $E^{b}_{2}$ and an antibonding one with its energy above $\eps_{2}$.  Within our convention, we therefore need a negative crystal field $\Delta_{const}$ to move the energy of the antibonding state down to $\eps_{1}$, and thus get back to the center of the high-spin region where the energy levels are degenerate. The bend of the high-spin region is simply due to the fact that, upon increasing $V_{1}$, the energy level of the antibonding $m=1$ orbital-ligand state is pushed up, therefore reducing the energy difference to the corresponding $m=2$ orbital-ligand state. \\
\begin{figure}[h]
\includegraphics[width=0.48\textwidth]{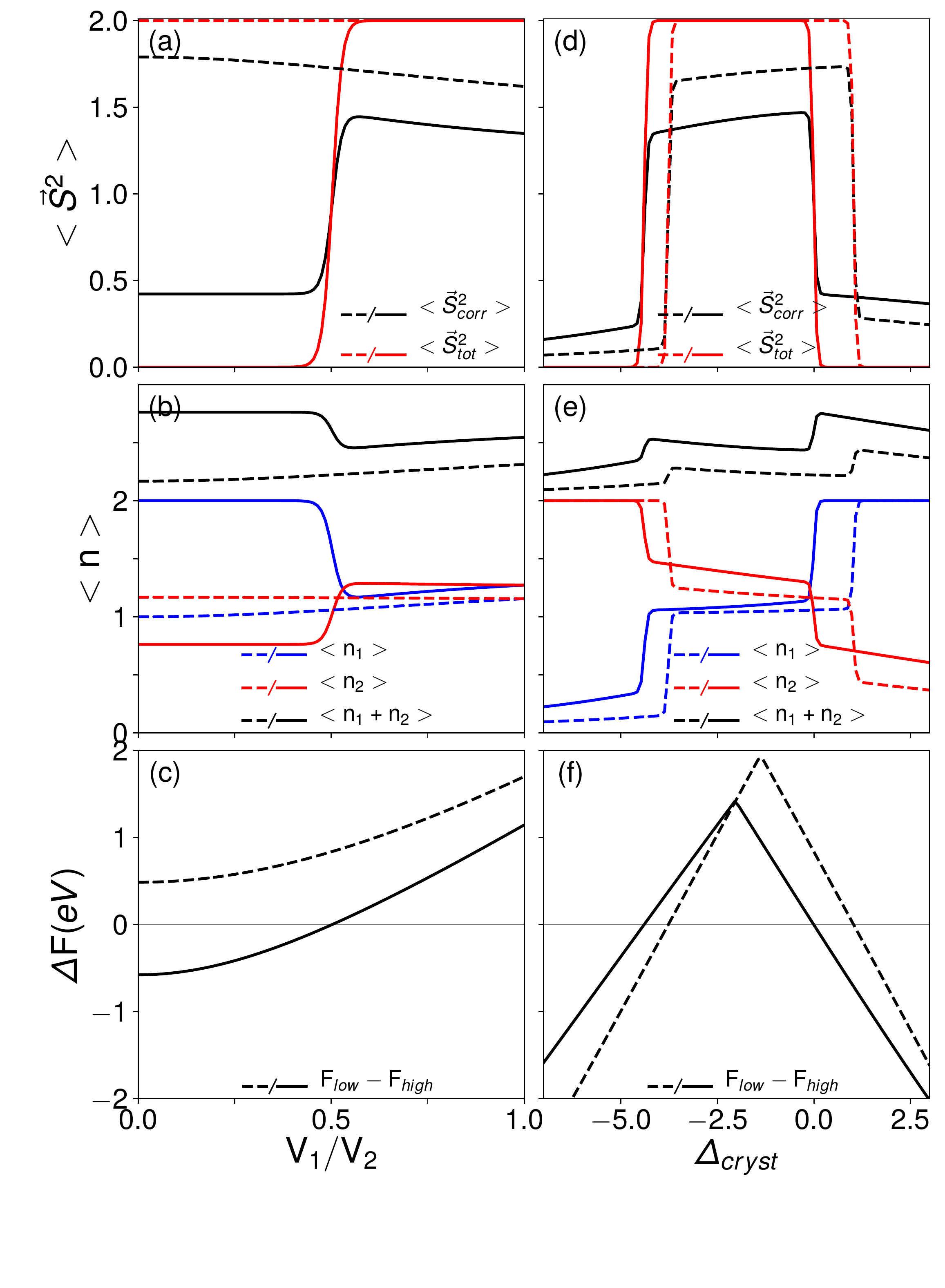}
\caption{Spin moment $\Braket{\vec{S}^2}$ (panels (a) and (d)), occupations $n$ of the correlated orbitals (panels (b) and (e)) and the difference in free energy between high and low-spin states (panels (c) and (f)) along the cuts indicated in Fig \ref{fig:generic_phasediagram} for $\epsilon_{H}(N=3)$ (solid lines) and $\epsilon_{H}(N=2)$ (dashed lines). Left side: Results along the cut at constant $\Delta_{cryst}=0$. Right side: Results along the cut with constant $V_{1}=V_{2}/2 = \sqrt{2}eV$.}\label{fig:equal_V1_CF_cut}
\end{figure}
%
\blue{
To get insight into the physics of the SCO, we look at the transition along two different paths in parameter space, marked as horizontal and vertical dashed lines in Fig. \ref{fig:generic_phasediagram}, respectively. The results are presented in Fig. \ref{fig:equal_V1_CF_cut}. Panels (a)-(c) show the results for the hybridization driven SCO, with vanishing crystal field $\Delta_{cryst}=0$; panels (d)-(f) explore the crystal-field driven SCO, with constant $V_{1} = V_{2}/2$.  Furthermore, we differentiate between results obtained for $\epsilon_{H}(N=3)$ (solid lines) and those for $\epsilon_{H}(N=2)$ (dashed lines)}. \\
The upper panels (a) and (d) show the spin moment along the cuts marked as dashed lines in Fig. \ref{fig:generic_phasediagram}. Red lines denote the spin moment of the full system $\Braket{\vec{S}_{tot}^{2}}$, black lines correspond to the correlated orbitals $\Braket{\vec{S}_{corr}^{2}}$ only. Panel (d) shows that a crystal-field driven SCO can be achieved for both values of $\epsilon_{H}$ under consideration; changing its numerical value merely leads to a shift of the transition points. 
Panel (a), however, makes clear that a hybridization driven SCO cannot be realized for $\epsilon_{H}(N=2)$ -- for which we remain in the high-spin regime for all $V_{1}$ under consideration -- but rather requires a bigger energy shift, as given by $\epsilon_{H}(N=3)$. 
In general, a comparison between the black and red  lines reveals that only the spin moment of the full system $\Braket{\vec{S}_{tot}^{2}}$ exhibits a clear low-spin to high-spin transition from $S=0$ to $S=1$. However, it is also the molecular spin moment, that is of experimental interest. \\
In the high-spin regime, one would expect the electrons to be (more or less) equally distributed among the two orbitals, while the low-spin regimes should be characterized by strong orbital polarization. Such behavior is indeed reflected in the middle panels (b) and (e) of Fig. \ref{fig:equal_V1_CF_cut}, which show the occupations of the different orbitals along the dashed lines drawn in Fig.\ref{fig:generic_phasediagram}. The roughly constant features of the dashed lines in panel (b) again witness the fact that no hybridization driven SCO is found for $\epsilon_{H}(N=2)$. Looking at the overall occupation $\Braket{n_{1}+n_{2}}$ in \blue{panel (e)}, one sees a ``staircase'' like behavior when changing the configuration from low to high-spin and back to low-spin. This can be understood as a consequence of the different hybridization strengths $V_{1}$/$V_{2}$, resulting in a different charge transfer to the correlated orbitals. \\
In the lower panels (c) and (f) of Fig. \ref{fig:equal_V1_CF_cut} we see the difference in free energy $F = E - TS$  between the lowest-lying (in terms of their energy) high/low-spin eigenstates $\Delta F = F[low] - F[high]$, as they are calculated along the cuts in Fig. \ref{fig:generic_phasediagram}. The point where this difference is zero marks the phase transition. The entropy is calculated as the \emph{ entropy} corresponding to the degeneracy of the eigenstate; while the high-spin state is two-fold degenerate with $S=\ln(2)$, the low-spin state is non-degenerate with $S=\ln(1)=0$.  Since $\beta=40$, this yields $TS \approx 0.017 eV$ for the high-spin state, which means that in our case the difference between energy and free energy is rather small. Comparison of the solid and dashed lines in panel (c) and (f) illustrates how $\epsilon_{H}$ shifts the energy difference between the high-spin and low-spin states, therefore underlining the different electron occupations of the correlated orbitals in the two regimes. 
\par
\emph{Conclusion.} The results of this model study indicate, that a purely hybridization-induced ($\Delta_{cryst}=0$) SCO cannot be accomplished in the naive scenario considering a static Hartree shift $\epsilon_{H}(N=2)$, corresponding to a half-filled e$_{g}$ manifold. However, as it can be seen from panels (b) and (e) of Fig. \ref{fig:equal_V1_CF_cut}, this assumption underestimates the actual average filling of the correlated orbitals. On the other hand, a SCO is found when considering a larger energy shift $\epsilon_{H}(N=3)$. This leads to the conclusion that the hybridization driven SCO, in the molecular setup under consideration, is intimately related to a charge-transfer from the ligands to the correlated orbitals. 
However, from Fig. \ref{fig:equal_V1_CF_cut} (b) and (e), it is clear that an energy correction corresponding to fixed electron numbers is at odds with the changing average occupations presented in these plots. 
In the following, when using parameters from realistic molecular structures, we shall improve on this inconsistency by applying a self-consistent double-counting scheme. 


%

\section{Spin-crossover in Ni-TPP}\label{sec:real_system}
Beyond the generic aspect, the applicability of our model to specific molecular complexes requires a precise \bblue{determination} of the model parameters. To this end, we perform density functional theory calculations for \blue{specific} molecular systems to derive \blue{the desired} model parameters in the flavor of the so-called DFT++ approach\cite{dft++,sumprb}. With this combined approach, in the following, we study the possibility of realizing a  spin-crossover in Ni-TPP isomers. 

\subsection{The system: Ni-TPP}
\begin{figure}[h!]
\begin{center}
\includegraphics[width=0.5\textwidth ]{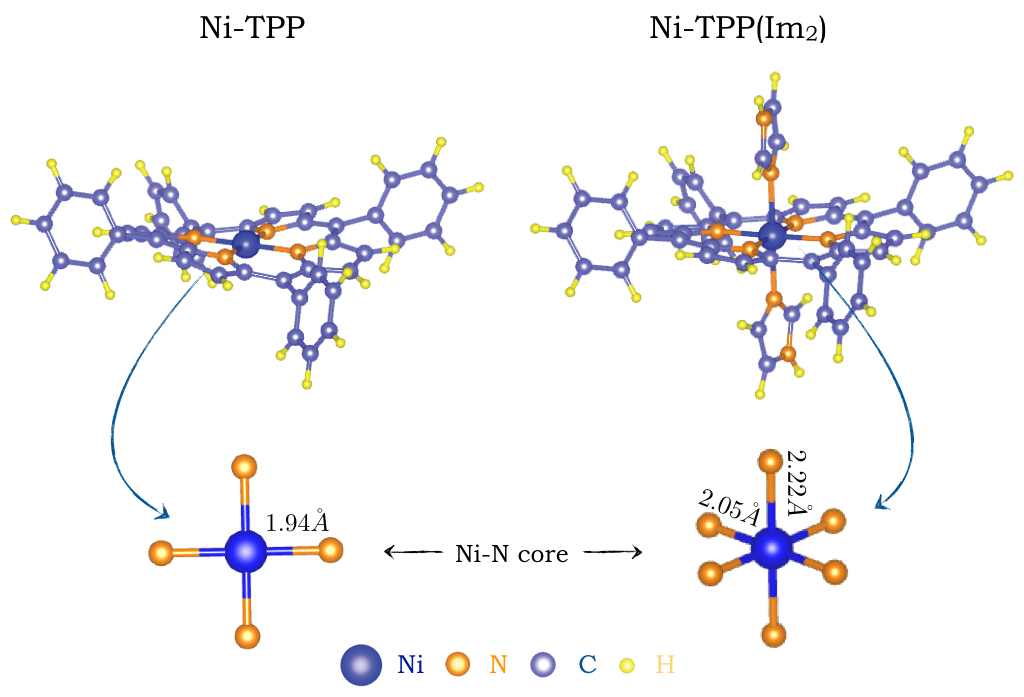}
\end{center}
\caption{\label{struct} The relaxed structures of gas-phase Ni-TPP and Ni-TPP (Im$_2$) molecules. The Ni-N cores of both the molecules are shown in the lower panel with axial Ni-N bond lengths.}
\end{figure}

\sbb{The Ni-porphyrin molecule contains a Ni$^{2+}$ ion in a porphyrin macrocycle, which is often attached to peripheral substituents, such as alkyl or aryl groups. The spin state of the molecule depends on the co-ordination of the central Ni-ion. The $3d$ orbital degeneracy of the Ni-ion is lifted by the ligand field provided by the organic ligands. 
In the four-fold coordinated Ni-porphyrin, the central Ni atom is exposed to a square-planar ligand field, leading to a low-spin state. The electronic structure \cite{theo_nip} for such a coordination suggests a fully filled \dzq orbital configuration, \blue{resulting in} a rather short Ni-N bond length\cite{same_structure,ruffle}. In heterosubstituated Ni-porphyrins, this short Ni-N bonds result in strong ruffling in the molecule, \blue{which} has a direct impact on the axial ligand affinity -- strong non-planarity \blue{was found\cite{ruffle} to} reduce the possibilities of ligand association. In Ni-porphyrin, however, the ruffling is moderate, \blue{allowing} an easier axial ligand association. \par
In Fig.\ref{struct} we present the relaxed structures of the four coordinated Ni-TPP (left) molecule and the six coordinated Ni-TPP (Im$_2$) (right) molecule with axial imidazole (Im) ligands. \blue{In both cases, one can observe a certain non-planarity, which is more strongly pronounced} in Ni-TPP than in Ni-TPP (Im$_2$). The Ni-N cores of  both molecules are presented in the lower panel of Fig.\ref{struct}. The Ni-N bond length in Ni-TPP is 1.94 \AA, which is 0.11 \AA~shorter than that in Ni-TPP(Im$_2$); in agreement with the experimental findings\cite{ruffle}. With a length of 2.22 \AA, the Ni-Im axial bond distance is much larger. The axial bond formation in Ni-TPP(Im$_2$), and the consequent expansion of the Ni-N core yields a weaker ligand field, and a high-spin state is stabilized.
Hence the two isomeric structures are characterized by two different, stable spin states and  thus constitute an ideal candidate for co-ordination induced spin-crossover (CISCO)\cite{cisco}.} 

\blue{Previous experiments demonstrated the possibility of a controlled CISCO in various molecules, including nickel porphyrin\cite{cisco,mri,room_temp_science,room_temp_switch}. In regard to potential applications in spintronic devices, however, it is more practical to achieve spin-state switching without changing the co-ordination number. Instead, we pursue a different approach, considering the six-fold coordinated Ni-TPP molecule with imidazole(Im) axial ligands under mechanical strain on the ligands. Upon ``stretching'' the axial ligands, one can modify the ligand field, possibly leading to a spin state transition. At a sufficiently  high strain, the axial ligands may dissociate, leaving a four coordinated Ni-TPP isomer.}


\subsection{Methods}
\subsubsection{DFT calculations }
The electronic structure of the free Ni-TPP isomers is calculated within density functional theory by using the Vienna ab-initio simulation package (VASP)\cite{vasp}. We use a plane wave projector augmented wave basis with the Perdew-Burke-Ernzerhof (PBE) -generalized gradient approximation of the exchange-correlation potential. To treat the isolated molecules we consider a  30x30x30 \AA$^3$ simulation cell which yields a minimum separation of 17.58 \AA~between the  molecule and its periodic image. The internal atomic positions are relaxed until the Hellman Feynman forces are minimized below 0.01 eV/\AA. In all our calculations, we used a  plane-wave energy cut-off of 400 eV. For the relaxation of the \st{molecules}\bblue{Ni-TPP, Ni-TPP (Im$_{2}$) molecules, as well as for the Ni-TPP (Im$_{2}$) with stretched imidazole ligands}, we employed the DFT+U formalism with Coulomb parameter F$_{0}$ = 4 eV and exchange parameter J$_{avg}$ = 1 eV, to account for the narrow Ni-$3d$ states. These interaction parameters correspond to the values chosen in the model calculations before Sec. \ref{sec:generic_results}, where we took the values corresponding to the e$_{g}$ manifold. The parameters for our model (Eq. \eqref{eq:the_model}), however, were retrieved from non-magnetic DFT calculations sans ``+U'' using the relaxed structures.  
 
 \subsubsection {Extraction of parameters from DFT calculations}
\par In order to employ our model Hamiltonian (Eq.~\ref{eq:the_model}) to realistic systems, we need to extract the correlated orbital onsite energies ($\epsilon_{m}$), as well as the hybridization-strength ($V_{m}$) and the ligand energies ($E^b_m$), as described in section \ref{model}. The latter two of these parameters enter the energy-dependent hybridization function
\begin{equation}\label{eq:hyb}
\Delta(\omega)_{i,j}=\sum_{m}\frac{V_{im}V_{mj}}{\omega+i\delta-E^b_m} \text{ .}
\end{equation}
In order to calculate the hybridization functions, the Kohn-Sham Green's function G$_{KS}$ is calculated from the Lehmann representation using
\begin{equation}
G_{KS}(\omega)=\sum_{nk}\frac{\ket{\psi_{nk}}\bra{\psi_{nk}}}{\omega+i\delta-\epsilon_{nk}},
\end{equation}
where $\psi_{nk}$'s and $\epsilon_{nk}$'s are the Kohn-Sham eigenstates and eigenvalues for band $n$ and reciprocal space point $k$. 

\bblue{In order to obtain the Green's function describing the local dynamics
of the correlated Ni-3d shell from the full Green's function of the
system, a projection of the Green's function onto the Ni-3d space is
performed. In practice, we are using the PAW implementation of VASP
\cite{vasp}, and a natural choice for defining the correlated local
space is given by the set of local orbitals $\chi_m$ that are atomic
Ni-3d wave functions within the augmentation sphere of the PAW method
and vanish outside of it.\\
The projectors $P^{m}_{nk}=\braket{\chi_{m} \mid \psi_{nk}}$ that describe the contributions of
the local orbitals to the Kohn-Sham eigenfunctions are normalized
using the overlap operators
\begin{align}
O_{mm'}(k)=\sum_nP^{m}_{nk}(P^{m'}_{nk})^* \text{ ,}
\end{align}
such that
\begin{align}
\tilde{P}^{m}_{nk}=\sum_{m'}[O(k)]^{-1/2}P^{m'}_{nk}\text{ .}
\end{align}
In this language, the local Ni-3d Green's function reads
\begin{equation}
             G^{mm^{'}}_{imp}(\omega)=\sum_{nk}\frac{\tilde{P}^{m}_{nk}{(\tilde{P}^{m'}_{nk}})^{*}}{\omega+i\delta-\epsilon_{nk}} \text{ .}
\end{equation}
}
Finally, the hybridization function is calculated from the local impurity Green's function by considering the expression
 \begin{equation}
            G^{-1}_{imp}(\omega)=\omega+i\delta-\epsilon-\Delta(\omega).
\end{equation}
In the above expression, \st{$G_{imp}$ is the projected Green's function in local orbitals and} $\Delta$ and $\epsilon$ are the hybridization function and the onsite energies of the \st{local} Ni-3d orbitals, respectively. 



\subsubsection{Double-counting}\label{sec:methods_double_counting}
In the context of realistic calculations, the quantity $\epsilon_{H}$, whose value has been kept constant during all calculations performed in Sec. \ref{sec:Results_onlymodel}, is identified with the double-counting potential
\begin{align}\label{eq:DC_potential1}
\epsilon_{H} \equiv \frac{\delta E^{dc}[\{\bar{n}^{\sigma}\}]}{\delta \bar{n}^{\sigma}} \text{ .}
\end{align} 
Here, we adopted the fully localized limit (FLL)\cite{Sawatzky_DC, Lichtenstein_LDA+U} approach, which can be considered appropriate for molecular systems. In this case, the energy correction becomes a function of the filling $N = \sum_{\si}N^{\si} = \sum_{\si}\Braket{n_{1\si}+n_{2\si}}$ of the correlated subspace and reads
\begin{align}\label{eq:dc_potential}
\frac{\delta E^{dc}[\{\bar{n}^{\si}\}]}{\delta \bar{n}^{\si}} = U_{avg}(N-\frac{1}{2}) - J_{avg} (N^{\si}-\frac{1}{2})
\end{align}
 (for details see Appendix \ref{sec:double_counting}). To avoid the inconsistency outlined at the end of Sec. \ref{sec:Results_onlymodel}, this double-counting correction is calculated in a charge self-consistent manner.\par
Evaluating the double-counting potential \eqref{eq:DC_potential1} self-consistently corresponds to solving an equation for the variable $\epsilon_{H}$. This equation can have multiple solutions, among which we chose the one with the lowest energy as the physical solution.

\subsection{Results}
\subsubsection{DFT results and connection to the minimal model }
\sbb{In Fig. \ref{schem}, we present the relative orbital energies of the  Ni-$3d$ orbitals, for  the Ni-TPP (left) and the Ni-TPP(Im$_{2}$) (right) isomers.  
In Ni-TPP, the highest occupied molecular orbital (HOMO) and the lowest unoccupied molecular orbital (LUMO) are of \dzq and \dxq character, with an energy separation of  1.79 eV. The degenerate \dxz and \dyz orbitals are close to \dzq (the energy difference being 0.04 eV), which itself is separated from the lowest-lying \dxy orbital by about 1 eV. These energies are in qualitative agreement with those from previous theoretical studies\cite{theo_nip}. A small difference is expected due to the different descriptions of the DFT exchange-correlation potentials. 
\begin{figure}[h!]
\begin{center}
\includegraphics[width=0.5\textwidth ]{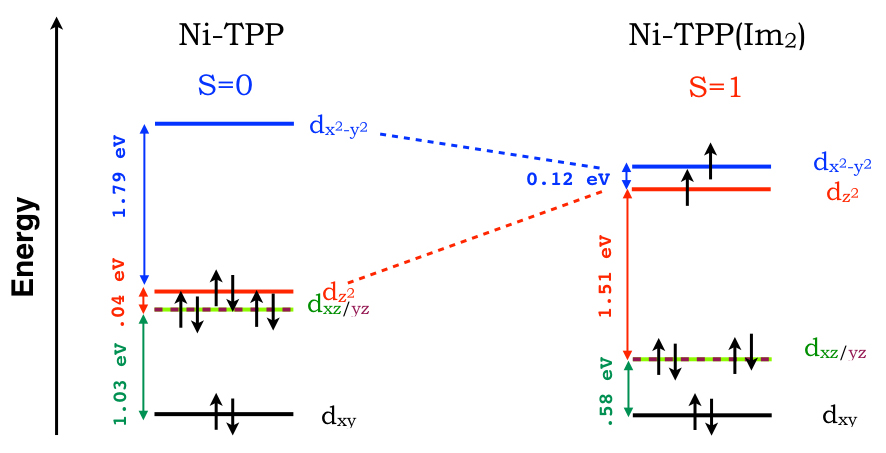}
\end{center}
\caption{\label{schem} Orbital energies of Ni-$3d$ orbitals in Ni-TPP (left) and Ni-TPP(Im$_2$) (right). The $t_{2g}$ orbitals are completely filled in both conformations, hence the magnetism is governed by the partially filled e$_{g}$ orbitals.}
\end{figure}

\blue{In Ni-TPP (Im$_2$),} the Ni-N bond length is extended by $\sim$6\%. Due to this core expansion, the energy of the \blue{\dxq orbital} is reduced, while the axial ligand bonding raises the orbital energy of \dzqX, reducing the corresponding energy separation to 0.12 eV. \blue{In accordance with the literature}, this results in a half-filled occupation of both orbitals, such that a high-spin triplet state is formed. The change of co-ordination has a crucial impact on the charge transfer between the porphyrin ring and the Ni ion. Within the DFT+U formalism (used to relax the molecular structures), the projected total charge on the Ni 3$d$ orbitals is 8.16 in the Ni-TPP (Im$_2$) molecule, while it is  8.4 in Ni-TPP. This charge transfer is enhanced in the many-body calculations, as will be described below.}\par
%
%
%
\begin{figure}[h!]
\begin{center}
\includegraphics[width=0.5\textwidth ]{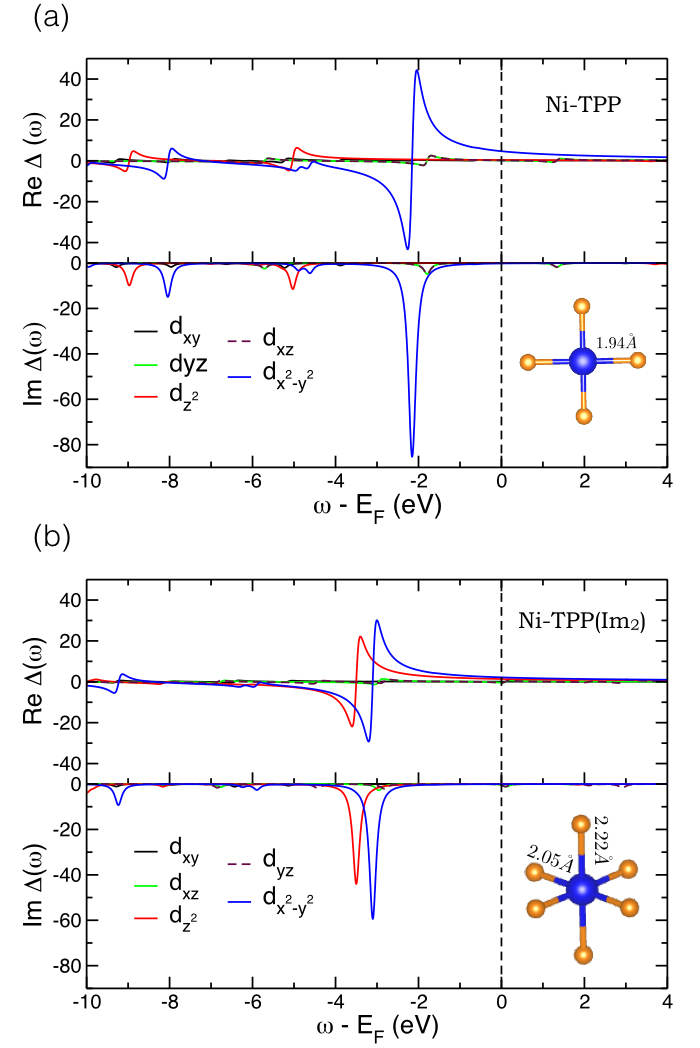}
\end{center}
\caption{\label{hyb} The real and imaginary parts of the energy-dependent hybridization functions  of Ni in Ni-TPP  (a) and  in Ni-TPP (Im$_2$) (b). A smearing parameter of 0.1 eV has been used for these plots for the sake of visualization.  In the inset, the corresponding co-ordination geometries of Ni atoms in Ni-TPP and Ni-TPP(Im$_2$) are shown, with axial Ni-N bond lengths.}
\end{figure}
In Fig. \ref{hyb}, we have plotted the real and imaginary parts of the hybridization functions $\Delta(\omega)$ of the Ni-$3d$ orbitals for the Ni-TPP (a) and Ni-TPP(Im$_2$) molecules (b). The corresponding Ni-N cores are presented in the in-set. In both molecules, $\Delta(\omega)$ has sharp peak structures, which is a signature of a confined system forming molecular orbitals. The hybridization function of Ni-TPP shows one dominant peak at -2.15 eV which corresponds to the \dxq orbital. The appearance of this peak results from the orbital overlap between the in-plane Ni-\dxq and the N-2$p$ orbitals. The other two peaks corresponding to the same orbital, appearing at -4.6 eV and -8.0 eV, respectively, are rather small and only have an insignificant effect on the low-energy physics of the system. For Ni-TPP-(Im$_2$), the dominant peaks in $\Delta(\omega)$ appear at -3.1 and -3.5 eV, corresponding to \dxq and \dzq orbitals, respectively. The reason behind the additional peak for the \dzq orbital is the axial bond formation between the \dzq and the N-$p_z$ orbital of the imidazole ligands. 
\begin{table}[h]
\caption{\blue{Parameters} as calculated from DFT calculations (using VASP) by projection onto localized orbitals. }
\centering
\begin{tabular}{l r|c|c}
& & \makecell{4-coordinated\\ Ni-TPP} &  \makecell{6-coordinated\\ Ni-TPP(Im$_2$)}\\
 \hline
 \hline
$\epsilon_{1}$ (d$_{z^{2}}$)& (eV) & -0.90 & -1.44\\
$\epsilon_{2}$ (d$_{x^{2}-y^{2}}$) & (eV) & -1.69 & -2.10\\ 
$|V_{1}|^{2}$ &(eV$^{2}$) & 0 & 4.39\\
$|V_{2}|^{2}$ &(eV$^{2}$) & 8.85 & 5.92\\
$E^{b}_{1}$ &(eV) & - & -3.5 \\
$E^{b}_{2}$ & (eV) & -2.15  & -3.1 \\
\hline
\end{tabular}
\label{tab:lda_parameters}
\end{table}
The intensity of the \dxq-hybridization peak is higher in Ni-TPP as compared to that in Ni-TPP-(Im$_2$); the former also appearing closer to the Fermi energy. 
This can be attributed to a stronger orbital overlap between \dxq and N-2$p$ orbitals in a shorter Ni-N bond.  A further inspection of the hybridization function for Ni-TPP(Im$_2$) reveals that the intensity of the \dzq peak is weaker as compared to that of \dxqX, with the latter one appearing closer to the Fermi energy. This is due to the fact that the Ni-N bonds with imidazole ligands are much larger (2.22 \AA) compared to that with the porphyrin ring (2.05 \AA), as shown in Fig.\ref{hyb}(b)(inset). The calculated values of the e$_g$ onsite energies, hybridization strengths and ligand energies for both Ni-TPP and Ni-TPP(Im$_2$) are summarized in TABLE \ref{tab:lda_parameters}. 
The hybridization of the Ni-$t_{2g}$ orbitals in both molecules is small, due to their non-bonding characters. In a d$^8$ (Ni$^{2+}$) configuration, the $t_{2g}$ orbitals, hence, remain completely filled, meaning that the relevant physics regarding the SCO is determined by the $e_{g}$ orbitals. \bblue{This confirms the choice of our model to describe the SCO.}

\subsubsection{Four-fold coordinated Ni-TPP}
The model Hamiltonian \eqref{eq:the_model}, provides a description of the physics of the four-fold coordinated Ni-TPP molecule, provided the parameters in the left column of Tab. \ref{tab:lda_parameters} are used. In this case, we find the ground state to be characterized by a low-spin moment, with $\Braket{S^2}_{corr} = 0.45$ for the correlated subspace and $\Braket{S^2}_{tot} = 0.0$. Keeping only states with a weight $>10^{-10}$, the ground state is spanned by only 4 Fock states, and can be written as 
\begin{align}\label{eq:Low_GS}
\begin{split}
\Ket{\mathbf{GS}}_{L} = 0.55\Ket{\ua\da,\ua}_{c}\Ket{\ua\da,\da}_{b} + 0.55\ket{\ua\da,\da}_{c}\ket{\ua\da,\ua}_{b} \\
 + 0.45\ket{\ua\da,0}_{c}\ket{\ua\da,\ua\da}_{b} + 0.44\ket{\ua\da, \ua\da}_{c}\ket{\ua\da,0}_{b} \text{ .}
\end{split}
\end{align}
In this notation, the subscript $c$ corresponds to the correlated orbitals, while $b$ designates the ligand states; the order is $\Ket{1,2}$. This ground state is characterized by a major charge transfer from the ligands to the correlated orbitals, such that the effective filling of the latter ones is close to three. The left panel of Fig. \ref{fig:energy_diagram} shows the energies of the lowest lying eigenstates, relative to the ground state, together with their spin $\Braket{\vec{S}_{tot}^{2}}$ and their degeneracy. \par
The states corresponding to the two lowest lying eigenenergies can be reproduced by a simplified Hamiltonian that considers only the Fock basis states that make up the ground state \eqref{eq:Low_GS}. A detailed discussion is found in Appendix \ref{sec:simplified_exptressions}, together with an explicit matrix representation of the corresponding Hamiltonian.

\subsubsection{Six-fold coordinated Ni-TPP(Im$_{2}$)}
\begin{figure}
\includegraphics[width=0.48\textwidth]{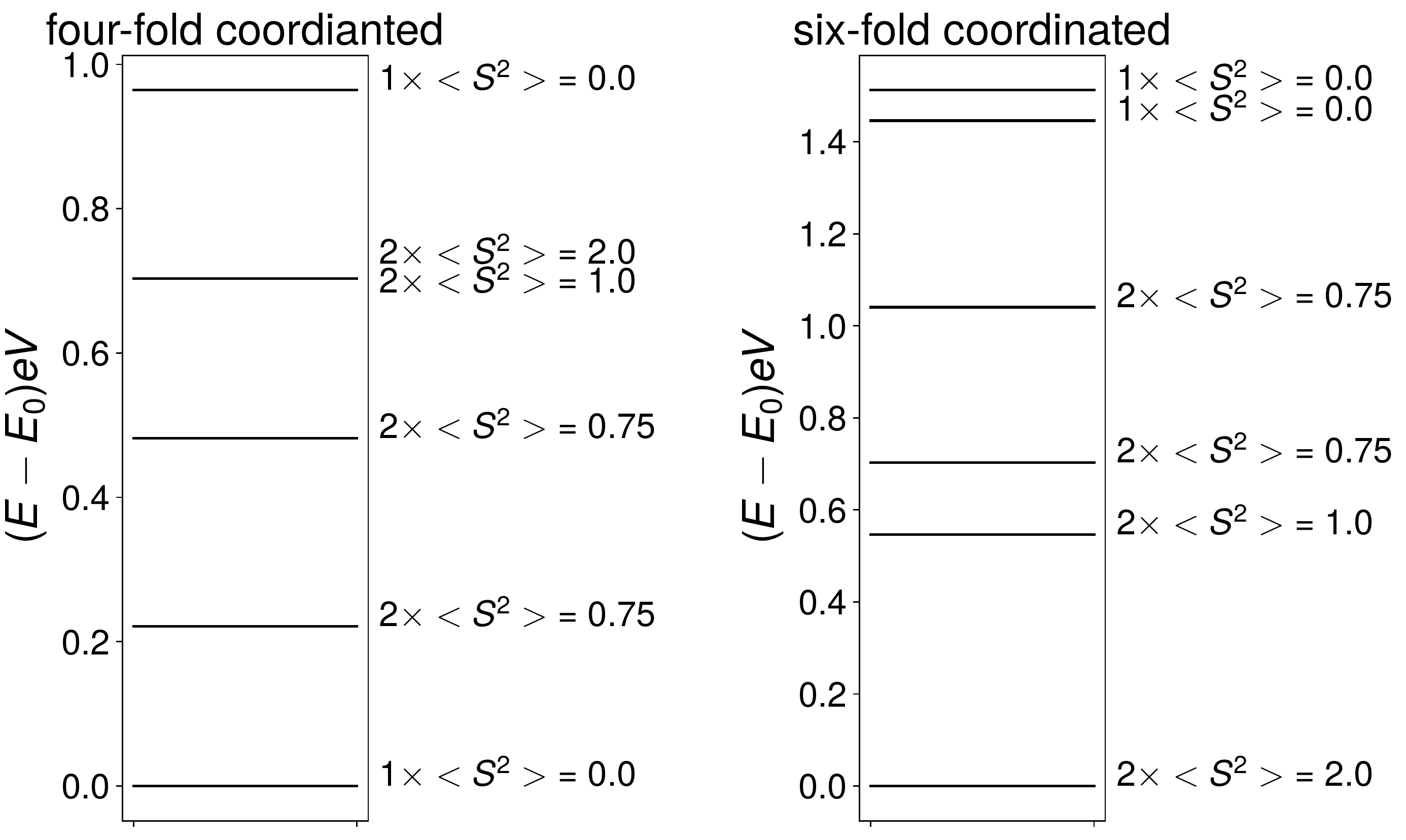}
\caption{Energy diagram of the infinitely stretched (Ni-TPP) molecule (left panel) and of the unstretched (Ni-TPP(Im$_{2}$)) molecule (right panel). All energy values are given relative to their ground-state; they are labeled according to their degeneracy and the corresponding spin value of the full system $\Braket{\vec{S}_{tot}^{2}}$, e.g. $2\times <S^{2}>= 2.0$ means that the energy is $2\times$ degenerate with a spin moment $\Braket{\vec{S}_{tot}^{2}}= 2.0$.}\label{fig:energy_diagram}
\end{figure}

We now turn to a discussion of the physics of the molecule with 
Ni-TPP(Im$_{2}$) configuration. Using the parameters of the right column of Tab. \ref{tab:lda_parameters}, the Hamiltonian \eqref{eq:the_model} has as a ground state a two-fold degenerate high-spin state with $\Braket{S^2}_{corr} = 1.54$  and $\Braket{S^2}_{tot} = 2.0$. The two states are spanned by 4 Fock states each, and are related by spin-flip symmetry 
\begin{align}\label{eq:High_GS}
\begin{split}
\Ket{\mathbf{GS}}_{H}^{(1)} = 0.81\Ket{\ua,\ua}_{c}\Ket{\ua\da,\ua\da}_{b} + 0.34\Ket{\ua\da,\ua}_{c}\Ket{\ua,\ua\da}_{b}\\
+ 0.46\Ket{\ua,\ua\da}_{c}\Ket{\ua\da,\ua}_{b} + 0.16\Ket{\ua\da,\ua\da}_{c}\Ket{\ua,\ua}_{b} \text{ ,}
\end{split}\\
\begin{split}
\Ket{\mathbf{GS}}_{H}^{(2)} = 0.81\Ket{\da,\da}_{c}\Ket{\ua\da,\ua\da}_{b} + 0.34\Ket{\ua\da,\da}_{c}\Ket{\da,\ua\da}_{b}\\
+ 0.46\Ket{\da,\ua\da}_{c}\Ket{\ua\da,\da}_{b} + 0.16\Ket{\ua\da,\ua\da}_{c}\Ket{\da,\da}_{b} \text{ .}
\end{split}
\end{align}
The filling of the correlated orbitals is $\Braket{n_{1}+n_{2}} = 2.38$, that is much closer to half filling then in the low-spin configuration. However, the charge transfer from the ligands is still considerable. The energies of the eigenstates, relative to the ground state, as well as the corresponding degeneracies and spin states can be found in the right panel of Fig. \ref{fig:energy_diagram}.\\
Again, the two lowest lying, non-degenerate energy levels are faithfully described by a simplified Hamiltonian, that acts on a truncated Hilbert space, spanned by the Fock basis states that make up the ground states \eqref{eq:High_GS}. As before, we refer to Appendix \ref{sec:simplified_exptressions} for a more detailed discussion.

\subsubsection{Spin-crossover: Strain on the axial ligands}

\begin{figure*}[t]
\hspace*{0cm}
\includegraphics[width=1.\textwidth]{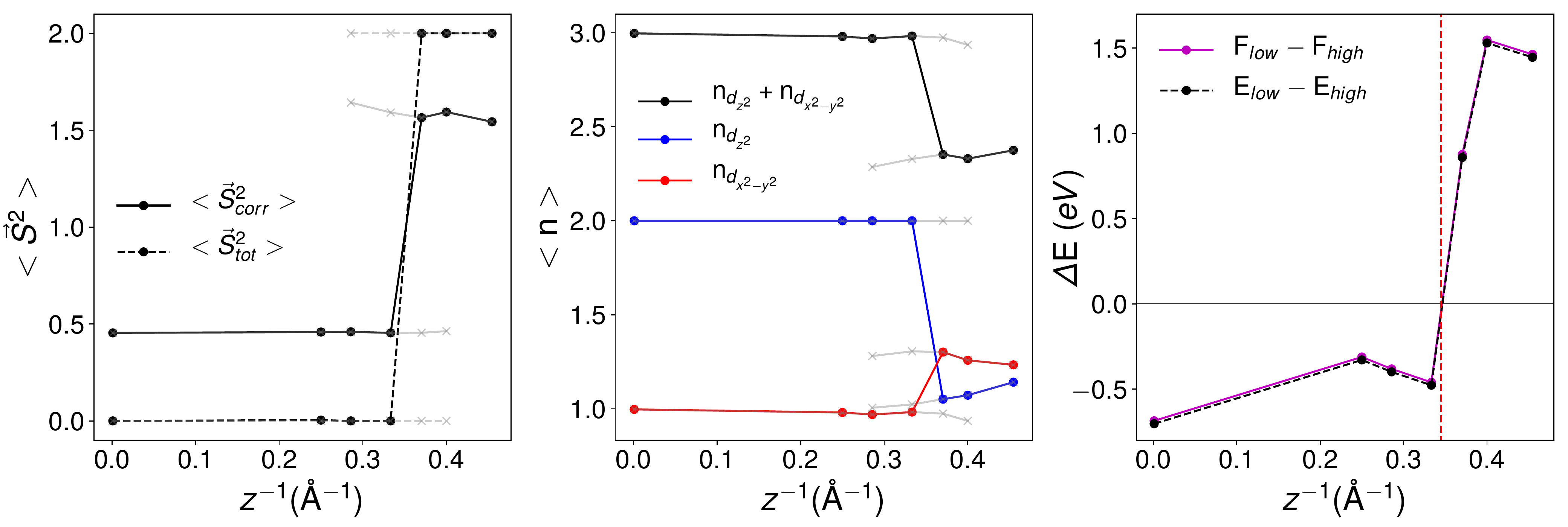}
\caption{Spin moment $\Braket{S^{2}}$ (left panel) , correlated orbital occupations $n$ (middle panel) and the difference in free energy and energy between the lowest lying low and high-spin states (right panel), as a function of the inverse \bblue{axial} bond length $z^{-1}$ between Ni and the imidazole ligands. Grey lines correspond to results obtained from analytic expressions, valid in the different regimes (see Appendix \ref{sec:simplified_exptressions}); \bblue{ the red, dashed line in the last panel indicates the SCO at $z\approx 2.90$\AA}.}\label{fig:stretched2unstretched}
\end{figure*}
%



\bblue{We finally turn to the physics of the strain-induced spin-crossover. In order to model the stretching of the axial N ligands, we relaxed (again applying the DFT+U method) the Ni-TPP (Im$_{2}$) structures (as explained above) for various fixed bond lengths between the Ni sites and the imidazole ligands, successively increasing the distance. The model parameters were then extracted from DFT calculations performed on these structures, in the same way as described above. Their values are listed in TABLE \ref{tab:lda_parameters2}. \par

\begin{table}[h]
\caption{\bblue{Parameters for the stretched Ni-TPP(Im$_{2}$) structure for different axial Ni-ligand bond lengths $z$, as calculated from DFT calculations (using VASP) by projection onto localized orbitals. }}
\centering
\begin{tabular}{l r|c|c|c|c|c|c}
z &(\AA)& 2.22$^{*}$& 2.50 & 2.70 & 3.00 & 3.50 & 4.00\\
 \hline
 \hline
$\epsilon_{1}$ (d$_{z^{2}}$) &(eV)& -1.44 & -0.98 & -1.20& -1.14 & -0.94 & -0.92\\
$\epsilon_{2}$ (d$_{x^{2}-y^{2}}$) &(eV)&  -2.10 & -2.27 & -2.44 & -2.19 & -1.92& -2.39\\ 
$|V_{1}|^{2}$ &(eV$^2$)&  4.39 & 2.04 & 1.18 & 0.44 & 0.10 & 0.00\\
$|V_{2}|^{2}$ &(eV$^2$)&  5.92 & 6.62 & 8.58 & 8.80 & 7.53 & 7.86\\
$E^{b}_{1}$ &(eV)&  -3.50 & -3.26& -3.18 & -2.80 & -2.46 & -  \\
$E^{b}_{2}$ &(eV)&  -3.10 & -3.06 & -3.03 & -2.73 & -2.54 & -2.39\\
\hline
\end{tabular}
\label{tab:lda_parameters2}
\end{table}

We then solved the model Hamiltonian \eqref{eq:the_model} for the corresponding parameters.
}
The results are shown in Fig. \ref{fig:stretched2unstretched}, which presents the variation of the spin moment  $\Braket{\vec{S}^{2}}$ (left panel), individual occupations of the correlated orbitals (middle panel), and the free energy of the spin configurations as a function of \bblue{the inverse axial Ni-N bond length $z^{-1}$. Presenting the observables as a function of the inverse \bblue{axial} bond length allows us to directly compare the results of the four-fold coordinated Ni-TPP molecule with those of the (stretched) Ni-TPP (Im$_{2}$) structures; the missing axial ligands correspond to $z_{\text{Ni-TPP}}=\infty$, such that $z^{-1}_{\text{Ni-TPP}}=0$.}  In the left panel of Fig. \ref{fig:stretched2unstretched}, solid and dashed lines correspond to the spin moment for the correlated orbitals $\Braket{\vec{S}_{corr}^{2}}$ and the whole molecule $\Braket{\vec{S}_{tot}^{2}}$, respectively. Our calculations predict a spin moment transition appearing at $z\approx 2.90$ \AA\, (from a linear interpolation of the free energies at $z=2.70$ \AA\,  and $z=3.00$ \AA). \par 
As the spin-state changes, a jump is observed in the occupations (middle panel). In the low-spin state, the total occupation of the correlated orbitals is $\sum_{\si}\Braket{n_{1\si}+n_{2\si}}$ $\approx$ 3 with \dzq orbital completely filled and \dxq orbital carrying $\approx$ 1 electron, owing to a strong hybridization with the N ligand. In the high-spin state, the Coulomb repulsion reduces the hybridization between \dxq and the molecular ligands, such that the occupation of both orbitals amounts to $\sim$1.2, reducing the total occupation to $\sum_{\si}\Braket{n_{1\si}+n_{2\si}}$ $\sim$ 2.4. \par
\blue{The grey lines in Fig. \ref{fig:stretched2unstretched} (left and middle panels) were obtained from the simplified models mentioned before (see Appendix \ref{sec:simplified_exptressions}), describing the physics in the two limiting cases.} The spin moment, as well as the occupations are almost perfectly reproduced within these models; they fail, however, to predict the spin moment transition, since the low-spin state remains energetically favored within the whole parameter range under consideration.\par
The right panel of Fig. \ref{fig:stretched2unstretched} shows the difference in free energy and energy between the lowest lying low and high-spin states as a function \bblue{of the inverse axial bond length $z^{-1}$}. Compared to Fig. \ref{fig:equal_V1_CF_cut} (c), one notes a jump in the region of the SCO, i.e. where the lines cross zero. This jump is due to the different occupations of the correlated orbitals corresponding to the different spin states. \par
\bblue{A similar high-spin low-spin transition as the one discussed here was studied in cis-dithiocyanatobis(1,10-phenanthroline)iron(II)
(Fe(phen)$_{2}$(NCS)$_{2}$) in Ref \onlinecite{C2CP40111H}. In that paper, the authors introduced an effective ``coordination number'' cn 
defined as a sum over a quantity characterizing the bond lengths. 
This number cn, which follows a thermal distribution function, was argued to determine the spin state of the molecule.
The temperature variation of the average cn is then a proxy for a temperature-induced 
spin-crossover.
In the light of our work, it is clear that also in that case the underlying mechanism is a change in the hybridization. The physical stimulus is however a different one. }

\section{summary}\label{sec:conclusion}
In summary, we have presented a minimal two-orbital model, that is able to capture the relevant physics of the spin-state transition in a given generic parameter space. We deduced a spin phase diagram which depicts the inter-dependence of hybridization and crystal field in order to bring in a spin-crossover. The model is complemented with parameters derived from DFT calculations providing a realistic scenario for the spin-crossover in Ni-TPP isomers. Our calculations show a robust low-spin state in Ni-TPP and a high-spin state in six-fold coordinated Ni-TPP(Im$_2$) \bblue{ accomplished by substantial Ni-ligand charge transfer. We then investigated the effect of mechanical strain by increasing the bond lengths with the  imidazole ligands of Ni-TPP(Im$_2$). 
The spin transition was found to appear upon moderately increasing the axial Ni-N bond to $z\approx 2.90$\AA, from its relaxed value of  $z\approx 2.22$\AA.} The ligand to metal charge transfer is enhanced as the molecular spin state changes from the high- to the low-spin state. We have discussed the importance of a charge self-consistent double-counting scheme in order to properly account for the metal-ligand charge transfer. Finally, our results suggest that a mechanical strain-induced SCO can be achieved in hexacoordinated Ni-TPP(Im$_2$). Such an effect would potentially allow to use it in \bblue{device set-ups based on} mechanically controlled or scanning tunneling microscopy break junctions.

  \bblue{
\section{Acknowledgments}
We thank La\"eticia Farinacci, Anna Galler and Hemlata Agarwala for helpful
discussions. This work was supported by the European Research Council (Consolidator Grant No. 617196 CORRELMAT) and supercomputing time 
at IDRIS/GENCI Orsay (Project No. t2019091393).
We thank the computer team at CPHT for support.
}

\appendix
\section{double-counting functional}\label{sec:double_counting}
When performing many-body calculations on top of results obtained from DFT, double-counting is an unavoidable problem. One, therefore, has to make sure to subtract contributions from interactions that were already taken into account within DFT. \\
A systematic way to deal with this redundancy is the inclusion of a double-counting functional
\begin{align}
\begin{split}
E^{DFT + X}[\rho^{\si}(\mathbf{r}), \{\bar{n}^{\si}\}] &= E^{DFT}[\rho^{\si}(\mathbf{r})] + E^{X}[\{\bar{n}^{\si}\}]\\
 &- E^{dc}[\{\bar{n}^{\si}\}] \text{ ,}
\end{split}
\end{align}
where $\rho^{\si}(\mathbf{r})$ is the DFT electron density and $\bar{n}^{\si}$ the orbital filling, both considering electrons of spin $\si$.\\
While there is a zoo of potential candidates, the fully localized limit  (FLL)\cite{Sawatzky_DC, Lichtenstein_LDA+U} approach can be considered the most suited for systems of molecular structure.\\
The FLL double-counting functional is given by
\begin{align}\label{eq:DC_lichtenstein}
E^{dc}[\{n^{\si}\}] = \frac{U_{avg}}{2}N(N-1) - \frac{J_{avg}}{2}\sum_{\si}N^{\si}(N^{\si}-1)\text{ ,}
\end{align}
which corresponds to the energy of the atomic configuration with degenerate orbitals. 
Here,  $N = \sum_{\si}N^{\si}$ and  $N^{\si} = \sum_{m} \Braket{n_{m\si}}$.
The variation of this functional with respect to the electron densities $n$ yields the double-counting potential
\begin{align}\label{eq:dc_potential}
\frac{\delta E^{dc}[\{\bar{n}^{\si}\}]}{\delta \bar{n}^{\si}} = U_{avg}(N-\frac{1}{2}) - J_{avg} (N^{\si}-\frac{1}{2})
\end{align}
which shifts the bare energy levels of the electrons.\\
For a full d-orbital shell, the averaged parameters can be expressed in terms of Slater parameters and are given by $U_{avg}= F_{0}$ and $J_{avg}= (F_{2} + F_{4})/14$ . However, since the model under consideration is restricted to the e$_{g}$ subspace only, this definition does not seem appropriate. In order to obtain more suitable expressions, we re-derive \blue{the averaged interaction parameters} for our e$_{g}$ system by considering the mean-field contributions to the interaction Hamiltonian 
\begin{align}
\begin{split}
H_{int}^{MF}[\bar{n}] &= \frac{1}{2}\sum_{m,m' \in e_{g},\si} U_{mm'}\bar{n}_{m\si}\bar{n}_{m'\bar{\si}}\\
&+ \frac{1}{2}\sum_{m\neq m'\in  e_{g}, \si} (U_{mm'}-J_{mm'})\bar{n}_{m\si}\bar{n}_{m'\si}\\
&= U_{avg}^{eg}N_{\ua}N_{\da} + \frac{1}{2}(U_{avg}^{eg}-J_{avg}^{eg})\sum_{\si}\frac{1}{2}N_{\si}^{2} \text{ .}
\end{split}
\end{align}
From the last equality we can directly read out the modified averaged interaction parameter
\begin{align}\label{eq:U_new_averaged}
U_{avg}^{e_{g}} &= \frac{1}{4}\sum_{m,m'\in e_{g}} U_{mm'} = U_{0} - J_{2} \text{ ,}
\end{align}
from which we further deduce the averaged Hund's coupling by demanding that 
\begin{align}
 J_{avg}^{ e_{g}} = U_{avg}^{ e_{g}}  -  \frac{1}{2}\sum_{m\neq m'\in  e_{g}}(U_{mm'}-J_{mm'}) = 2J_{2} \text{ .}
\end{align}
These are the modified parameters we used with functional \eqref{eq:DC_lichtenstein} to correct for double-counting.\\

\section{Simplified asymptotic models}\label{sec:simplified_exptressions}
In the limiting cases of the bi-pyramidal and (infinitely stretched) square planar configurations, corresponding to the parameter set (Tab. \ref{tab:lda_parameters}) obtained from our DFT calculations, the results obtained from Hamiltonian \eqref{eq:the_model} correspond to those obtained from two very simple models, which shall be described in the following.\\
\textbf{Low-spin case.} In the infinitely stretched case, the \dzq orbital does not hybridize any more with its ligand, therefore making the \dzq filling a good quantum number. The Coulomb repulsion acting on the \dxq orbital will therefore manifest itself as a mere shift of the \dxq bare energy. Due to the strong hybridization of the \dxq orbital with the ligand, the effective \dxq energy will lie above the one of the \dzq, justifying the  assumption that the \dzq orbital (and it's ligand) are completely filled. Since our system contains $6$ electrons, this leaves us with a strongly reduced Hilbert space, spanned by the $4$ states
\begin{align}
\begin{split}
&\Ket{\ua\da,\ua}_{c}\Ket{\ua\da,\da}_{b} \text{ , } \Ket{\ua\da,\da}_{c}\Ket{\ua\da,\ua}_{b} \text{ , }\\
&\Ket{\ua\da,0}_{c}\Ket{\ua\da,\ua\da}_{b} \text{ , } \Ket{\ua\da,\ua\da}_{c}\Ket{\ua\da,0}_{b} \text{ .}
\end{split}
\end{align}
Using these as our basis states, we can write the corresponding Hamiltonian as a matrix
\begin{widetext}

\begin{align}
H_{\text{low}} = \begin{pmatrix}
2U - 5J + \Delta/2 - \mu + E^{b}_{2} & 0 & -V_{2} & -V_{2}\\
0 & 2U - 5J + \Delta/2 - \mu + E^{b}_{2} & V_{2} & V_{2}\\
-V_{2} & V_{2} & 2 E^{b}_{2} & 0\\
-V_{2} & V_{2} & 0 & 5U - 10J + \Delta - 2\mu 
\end{pmatrix} \text{ ,}
\end{align}

\end{widetext}
which can be easily diagonalized.\\
\textbf{High-spin case.} In the bi-pyramidal configuration, Hund's coupling will strongly favour the states
\begin{align}
\Ket{\ua,\ua}_{c}\Ket{\ua\da,\ua\da}_{b} \text{ , } \Ket{\da,\da}_{c}\Ket{\ua\da,\ua\da}_{b} \text{ .}
\end{align}
Since the second state can be generated from the first one by applying a global spin-flip transformation (which, without any external magnetic field leaves the system invariant), it suffices to consider only the first one in the following. Due to the considerable hybridization $V_{1}$, $V_{2}$ with the ligands, we also have to consider the states
\begin{align}
 \Ket{\ua\da,\ua}_{c}\Ket{\ua,\ua\da}_{b} \text{ , }\Ket{\ua,\ua\da}_{c}\Ket{\ua\da,\ua}_{b} \text{ , } \Ket{\ua\da,\ua\da}_{c}\Ket{\ua,\ua}_{b} \text{ .}
\end{align}
Thus, our Hilbert space is again of dimension $4$, and we can write down the Hamiltonian as a matrix
\begin{widetext}
\begin{align}
\resizebox{1\hsize}{!}{$
H_{\text{high}} = \begin{pmatrix}
U - 3J - 2\mu + 2(E^{b}_{1} + E^{b}_{2}) & -V_{1} & -V_{2} & 0 \\
-V_{1} & 3U - 5J - \Delta/2 - 3\mu + E^{b}_{1} + 2 E^{b}_{2} & 0 & -V_{2}\\
-V_{2} & 0 & 3U - 5J + \Delta/2 - 3\mu + 2E^{b}_{1} +  E^{b}_{2} & -V_{1}\\
0 & -V_{2} & -V_{1} & 6U - 10J  - 4\mu + (E^{b}_{1} + E^{b}_{2})
\end{pmatrix} \text{ ,}
$}
\end{align}
\end{widetext}
which can be diagonalized easily. \par
Fig. \ref{fig:stretched2unstretched} shows the results from the simplified models as grey lines. Within the high-spin/low-spin regions, the results from the corresponding asymptotic expressions are very close to the curves from the full model. However, they do not reproduce SCO (as explained in the main text).

\nocite{*}
%

\end{document}